# Detecting Ideal Instagram Influencer Using Social Network Analysis

M.M.H Dihyat
Electronic Engineering and Computer Science (EECS)
Queen Mary University of London
London, United Kingdom
m.dihyat@se17.qmul.ac.uk

M.A Khan
Electronic Engineering and Computer Science (EECS)
Queen Mary University of London
London, United Kingdom
mehmood.khan@se17.qmul.ac.uk

K Malik
Electronic Engineering and Computer Science (EECS)
Queen Mary University of London
London, United Kingdom
k.malik@se17.qmul.ac.uk

B Imran
Electronic Engineering and Computer Science (EECS)
Queen Mary University of London
London, United Kingdom
b.s.imran@se17.qmul.ac.uk

*Abstract—* Social Media is a key aspect of modern society where people share their thoughts, views, feelings and sentiments. Over the last few years, the inflation in popularity of social media has resulted in a monumental increase in data. Users use this medium to express their thoughts, feelings, and opinions on a wide variety of subjects, including politics and celebrities. Social Media has thus evolved into a lucrative platform for companies to expand their scope and improve their prospects. The paper focuses on social network analysis (SNA) for a real-world online marketing strategy. The study contributes by comparing various centrality measures to identify the most central nodes in the network and uses a linear threshold model to understand the spreading behaviour of individual users. In conclusion, the paper correlates different centrality measures and spreading behaviour to identify the most influential user in the network.

## I. INTRODUCTION

Instagram is a popular social media platform where users are able to share photos, video and follow other users. Users such as Instagram influencers are able to take advantage of this platform for monetary opportunities. Instagram influencers are content creators who create communities around topics and niches by sharing their work on the platform. Instagram is a visual medium, so it works best for visually dynamic content. Lifestyle influencers and celebrities who share facets of their personal lives on the site, as well as content creators covering travel, photography, fashion, beauty, fitness, food, and other subjects, find success on Instagram. Instagram influencer marketing is now a $2.3 billion industry that has swept the social media landscape. Instagram users all over the world have started making money from paid posts and collaborative marketing campaigns in recent years. [1]

The paper concentrates on the social media marketing campaign conducted by a restaurant named Veganwich which is located in Manchester. Coopers, the founders of the restaurant have decided to use instagram influencers as part of their initiative to help promote their new vegan sandwich shop. Despite their lack of experience with social media for companies, the Coopers understand that using these online channels to spread the word about their menu to potential customers can be extremely helpful to their company. Thus, to attract the local community, they have decided to select a local influencer.

The aim of this project is to use social network analysis in order to produce a plan to make an effective influencer marketing strategy. This will provide a simple and intuitive network visualisation, as well as information about the network's configuration and distinguishing features. Our marketing plan would be backed up by a network-wide comparison of various centrality measures. An investigation will be conducted to find the benefits and drawbacks of the centrality measures when applied in the context of our research. This will be linked to influencer marketing and the structures that control how social marketing operates. Based on our analysis and produced influencer marketing strategy, suitable influencers would be picked for the Coopers to offer their sponsorship deal.

The paper consists of five further sections: Section 2 reviews recent studies on influencer marketing strategies and examines the project's numerous hypotheses. The configuration of the dataset used in the study, as well as the overall network structure, is discussed in Section 3. In Section 4, the network analysis and methodologies are detailed. In Section 5, the methodologies' results are discussed along with the derived recommendation based on the results. In conclusion, a concise summary will be given in section 6 with regards to the study

## II. RELATED WORK

A social media influencer is someone who has built a reputation based on their knowledge and expertise with the ability to influence others in the society. So how do seemingly normal individuals get the ability to influence others?

Studies have found that the more interactive a public persona is, the more likely it will generate higher affinity and trust [2]. The cues of interactivity and engagement include numbers of followers, followings, shares, likes and comments. These indices are more significant for non-mainstream celebrities because a "regular" person's high numbers of followers and likes can be attributed to this person's active engagement, openness to audiences and the popularity in the online community [3]. Traditional celebrities, on the other hand, with large numbers of followers and likes can be perceived as an extension of their offline popularity, regardless of their sociability or proactivity on social media.

As suggested in the word 'influencer', the aim of such individuals is to sell their sponsors' product. Customers tend to look for the attraction of the product which can be measured by the positive reviews or by the number of likes on social platforms. However, gaining the likes and positive review cannot be possible if the influencers do not possess the skills of bringing out the true values of the product. Instagram fashion influencers are perceived to be more authentic when their visual appearance and luxurious lifestyle match the symbolic value of luxury brands [4]. When people encounter a visual image of Instagram celebrity with a luxury product, they would form a positive attitude toward the featured brand if they identify more with the source [5].

Influencer marketing has been shown to be extremely successful in studies. For example, according to a 2016 survey, 40% of respondents purchased an item after seeing it used by an influencer on Twitter, YouTube, Instagram, and other social media platforms. According to the study, when Twitter users were exposed to both brand and influencer tweets about the product, their purchase intent increased by 5.2 times [6]

In August 2020, HelloFresh who are a German meal-kit company started an online marketing campaign by utilising micro-influencers (people who have 1,000 to 100,000 followers) on instagram. The brand partnered with 15 UK based micro-influencers to promote home cooking. To ensure the campaign reaches the mass audience, each influencer encouraged their followers to join the campaign by using the tag #RefreshWithHelloFresh. This campaign led to an increase of 274% increase in brand impressions as well creation of 325% creation of extra content. All these led to a significant increase in brand recognition for HelloFresh [7]

Finding individual superspreaders and a group of multiple influencers that can maximise global impact are two separate topics in the problem of identifying influencers [8]. In this paper, we investigate the first problem, i.e., finding the individual superspreader [9]. Network theory allows us to visualise real world networks as graphs where individual users are denoted as nodes. Identifying influential nodes that contribute to faster and wider dissemination of information in real world networks is thus critical for a successful influencer marketing strategy.

In 2018, Rum et al used social network analysis to detect social media influencers in the health and beauty industry of Malaysia. As part of their study, they conducted a statistical analysis of three centrality measures: closeness centrality, betweenness centrality, and eigenvector centrality. They also looked at the correlation between the centrality variables and discovered that eigenvector centrality was strongly correlated with both betweenness and closeness centrality and was thus the best measure to identify influential nodes in a network. As a result of their findings, they came to the following conclusion - "the more important the users in the network, the more central and closer the node is to the other users". [10]

According to Bryungjoon et al, the location of a node in a network has an impact on its spreading ability. They compared nodes found in the core to those found in the periphery. They discovered that a node in the network's core can have a much greater effect on spreading than a node in the periphery. They identified the location of a node in the network using k-shell decomposition and analysed its spreading behaviour using a SIR (Susceptible, Infected, and Recovered) epidemic model [8]

### III. DATASET AND NETWORK PREPARATION

The dataset was collected through Instagram and it is collated with Instagram users that reside in Manchester with a recent history (previous six months) of using the hashtag #vegan in their posts. The dataset is visualised in the form of a two-column Comma-Separated Values (CSV) file.

| i | j |
|---|---|
| 1 | 2 |
| 1 | 3 |
| 4 | 5 |
| 6 | 7 |
| 8 | 9 |
| 10 | 11 |
| 12 | 13 |
| 12 | 14 |
| 12 | 15 |
| 12 | 16 |
| 12 | 17 |
| 18 | 19 |

*Table 1: Visualisation of the Cooper's Instagram Network*

#### A. Data Representation

The two columns "i" and "j" represent two nodes, where the node in column "j" is being followed by node "i". The dataset represents a directed graph due to the nature of Instagram. Where one has to follow another to see their posts. The total number of Instagram users (nodes) given in the dataset are 874 and the combined number of connections (edges) between nodes are 1853.

#### B. Softwares Used

The dataset was passed through an open-source network analysis and visualisation software package called Gephi. The dataset was converted into a network model defining the connections between the nodes and a visualisation of the communities within the network. It helped in calculating the basic statistics and visualising the structure of the network. Furthermore, the NetworkX library was used for more complex research. NetworkX is a Python package for creating, manipulating, and studying complex networks' structure, dynamics, and functions.

## C. Network Visualisation

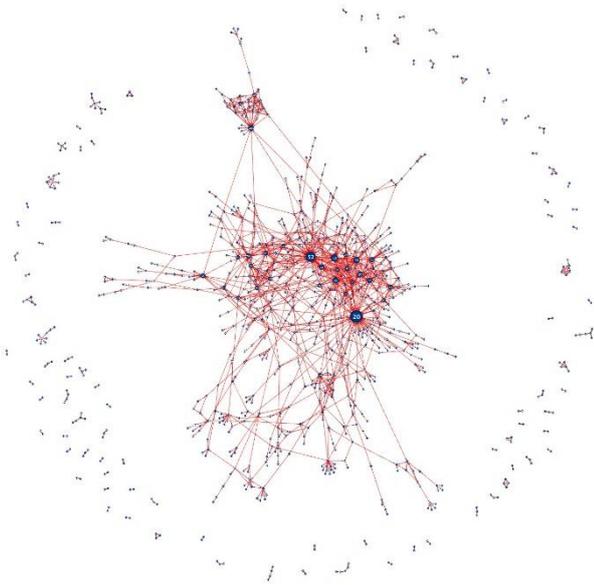

*Figure 1: Network Visualisation*

The force atlas2 layout was used to create the network visualisation. This layout was chosen because it clearly depicts the network's core and periphery. The nodes are shown in a dark blue colour, and their size is determined by the number of in-degrees they contain, while the edges are shown in a red colour.

As we can clearly see, the network consists of a highly connected central core and many disconnected components present at the outskirts of the network. In context of our problem, it is impossible to reach the users present in the disconnected components, unless we pick influencers present in specific components. Hence, we are only considering the central core of the network for finding influencers. The total number of nodes in the network's central core is 610 which means that 69.79% of the total nodes in the network are present in the central core while the total number of edges is 1604. This indicates that approximately 69.79% of the users (nodes) are in the network's central core.

## D. Basic Statistics

| Network | Number of nodes | Number of edges | Average path length | Average clustering coefficient | Diameter | Connected Components |
|---|---|---|---|---|---|---|
| Cooper instagram network | 874 | 1853 | 4.69 | 0.13 | 15 | 95 |
| $R_1$ | 874 | 19059 | 2.48 | 0.02 | 10 | 1 |
| $R_2$ | 874 | 37861 | 2.049 | 0.05 | 8 | 1 |

*Table 2: Basic statistics*

The above table shows the basic statistics of the instagram network and two random networks $R_1$ and $R_2$. The random networks are generated with the same number of nodes as the instagram network, and with a 0.05 and 0.10 rewiring probability, respectively. The Cooper instagram network's average path length and average clustering coefficient are 4.69 and 0.13, respectively, for $R_1$ the values are 2.48 and 0.02 and finally for $R_2$, the values are respectively 2.049 and 0.05. All three networks have similar average path lengths, but the clustering coefficients between the random networks and the Cooper network are vastly different. This demonstrates that the Cooper Instagram network is a small world network.

## IV. NETWORK ANALYSIS METHODOLOGY

Our analysis makes use of three centrality measures to identify the most central node in the network: Degree, Betweenness and Eigenvector centrality. In addition, a spreading capability of the node is calculated through a cascade threshold to calculate the adoption of an action or idea based on the influence. This paper illustrates the correlation of spreading behaviour with different centrality measures and identifies the optimal influencer in the graph. In this section, we at first look at the different centrality measures and define the top ten nodes for each one. Then, using a linear threshold model, we observe the information spreading behaviour for each node. Finally, the nodes are ranked based on their ability to disseminate information.

### A. Centrality Measures

In graph theory, centrality is characterised as a measure of a node's importance within a graph. Since social networks are inextricably linked to graph theory, and the issue of identifying the most prominent users in a social network is, at the end of the day, a measure of significance, centrality measures have received the most attention in the early years of influencer recognition research [11]

The most basic and historic centrality measure is called "**Degree** centrality" which is defined as the number of connections of a node. In a directed network, each node has two measures which are called indegree and outdegree. In terms of an instagram user, we can define indegree as the number of followers the user has and outdegree as the number people the user follows. While degree centrality can help us identify nodes with the most number of followers, it cannot perceive the true influence of the node in a network.

In certain cases, the influential nodes may have fewer followers, but still may distribute knowledge efficiently. **Betweenness** Centrality is conceived as a measure for quantifying the control of a human on the communication between other humans in a social network and thus a highly useful measure in influencer detection [12].

Another centrality measure that comes into consideration is the Eigenvector centrality. Eigenvector centrality is a metric of a node's significance that considers its neighbours' importance. For example, "a node with 300 relatively unpopular friends on Facebook would have lower eigenvector centrality than someone with 300 very popular friends (like Barack Obama)" [13].

We determined the value of the three centrality measures for each node in the network. We then chose the ten nodes with the highest values for each centrality measure. These nodes were chosen to better understand the relationship between the most central nodes and their spreading behaviour.

### B. Spreading behaviour

Predicting the spreading capacity of a node in a network can be done using a variety of methods and models. Bryungjoon et al., for example, used an SIR model with a wide range of infection probabilities [8]. A Linear Threshold model is used in the same way. It's a diffusion model in which each node has a threshold value $\Theta_i$ in the range [0, 1], and when the number of active neighbours of the node (i) exceeds the value $\Theta_i$, the node (i) becomes active and adopts the behaviour [11]. Moreover, this model also resembles social learning in real life. According to Salganik et al, "social influence exerts an important but counterintuitive effect on cultural market formation, generating collective behaviour that is reminiscent of information cascades in sequence of individuals making binary choices" [14]. The statement of Salganik et al is supported by Harvard Business Review (HBR) in which 19% of all US customers bought a product or service after seeing it recommended by a social media influencer in 2018 [15].

In our Linear Threshold model, we selected three different thresholds: 0.05, 0.1 and 0.2 to observe different spreading behaviours in the Cooper's instagram network and chose to run the model for 15 days to capture the efficiency of adoptions. Moreover, by default, the linear threshold model assumes that behaviour spreads from source to target. However, due to the nature of instagram, behaviour is actually passed in the opposite direction. As a result, we have inverted the graph, which means all the edge directions were reversed. This inverted graph was only used in the linear threshold model, not in the calculation of centrality measures.

### C. Ranking Nodes and Selecting the Optimal Node

For the given starting node (influencer), a threshold value and number of iterations (days), the Linear Threshold model simulates the spreading behaviour initiated from that node. In other words, the model observes how the behaviour spreads from a starting node. It calculates the proportion of nodes in the network reached in the given number of iterations (days).

There is a limit on how far a node can distribute information throughout the network. The node's scope does not increase after reaching the influence limit; it remains constant. As a result, the number of days a node takes to reach the influence limit can be used to determine its spreading capacity. Therefore, the spreading capability can be expressed as -

$$Spreading\ capability(n) = \frac{Proportion\ of\ nodes\ reached}{Number\ of\ days\ required\ to\ reach\ the\ limit}$$

Where n is the starting node, the number of days required to reach the limit > 0

Furthermore, the spreading capability of selected 30 nodes can be ranked after measuring their spreading behaviour. The node with the highest value would reach the highest number of nodes in the network in the shortest amount of time. As a result, the node with the highest spreading capacity value will be the optimal node. In the context of the problem, the most optimal influencer is one that can spread the news quickly and to a large number of users.

## V. RESULT AND DISCUSSIONS

### A. Calculating Centrality Measures

We use the in-degree of the nodes to measure the degree centrality since the relationship between influencers and followers is best expressed as in-degrees. The top ten nodes with the most indegrees are shown in the diagram below:

|   | Node | In-degree | Out-degree |
|---|------|-----------|------------|
| 0 | 20   | 59        | 59         |
| 1 | 52   | 46        | 36         |
| 2 | 53   | 28        | 22         |
| 3 | 36   | 23        | 21         |
| 4 | 32   | 22        | 22         |
| 5 | 185  | 20        | 22         |
| 6 | 37   | 20        | 16         |
| 7 | 261  | 17        | 14         |
| 8 | 51   | 17        | 18         |
| 9 | 44   | 17        | 23         |

*Table 3: Degree (In-degree) Centrality Measure*

Node 20 has the most indegrees in the network, as seen in the table above. However, since the centrality measure does not reflect the value of the node or how central it is, we are unable to determine if this node will affect the greatest number of users. In other words, an influencer's number of followers does not imply the influencer's spreading capability or whether it is the most central and significant member of the network. Thus, to understand which nodes can spread information further, we calculated the betweenness centrality as shown in the below table:

|   | Node | Betweenness | In-degree | Out-degree |
|---|------|-------------|-----------|------------|
| 0 | 20   | 0.17975     | 59        | 59         |
| 1 | 52   | 0.09510     | 46        | 36         |
| 2 | 185  | 0.05680     | 20        | 22         |
| 3 | 261  | 0.04447     | 17        | 14         |
| 4 | 110  | 0.04254     | 13        | 17         |
| 5 | 53   | 0.04210     | 28        | 22         |
| 6 | 415  | 0.03425     | 2         | 2          |
| 7 | 36   | 0.03265     | 23        | 21         |
| 8 | 310  | 0.02748     | 6         | 6          |
| 9 | 316  | 0.02421     | 8         | 15         |

*Table 4: Betweenness Centrality Measure*

In comparison to the other nodes, node 20 disseminates information more efficiently, as seen in the table above. Furthermore, we discovered that a large number of indegrees does not solely affect the efficient spread of information. Node 53, for example, has a greater number of 'in degrees' than node 185, but node 185 has a higher betweenness score. While betweenness centrality is a useful measure for identifying nodes that efficiently distribute information, it does not always enable us to identify network influencers. In table x, node 415, for example, has a high betweenness value as compared to node 36. However, upon closer inspection, we can see that node 36 has a far higher in-degree than node 415, indicating that node 36 is a more powerful influencer while node 415 acts as a bridge in the network. This illustrates that betweenness centrality measure solely is not a good measure to identify influencers.

Table 5 illustrates that node 20 has an eigenvector value of '0.39108' which is the highest among its peers, as well as a high betweenness value. Not all nodes with a high betweenness, on the other hand, have a high eigenvector value. Node 36, for example, has a low betweenness value of 0.03265 and a high eigenvector value of 0.26058. In terms of the Coopers network, this means that even though node 36 is situated near highly influential nodes, it however cannot spread information quickly among its followers.

| | Node | Eigenvector | In-degree | Out-degree |
|---|---|---|---|---|
| 0 | 20 | 0.39108 | 59 | 59 |
| 1 | 52 | 0.34206 | 46 | 36 |
| 2 | 36 | 0.26058 | 23 | 21 |
| 3 | 53 | 0.25556 | 28 | 22 |
| 4 | 32 | 0.25109 | 22 | 22 |
| 5 | 37 | 0.23278 | 20 | 16 |
| 6 | 46 | 0.21539 | 16 | 20 |
| 7 | 51 | 0.20250 | 17 | 18 |
| 8 | 33 | 0.19929 | 15 | 19 |
| 9 | 44 | 0.18445 | 17 | 23 |

*Table 5: Eigenvector Centrality measure*

As a result of the above comparisons, we can conclude that only choosing one centrality measure does not allow us to identify the optimal influencer in the network.

### B. Spreading Behaviour

We used four different thresholds in our Linear Threshold model: 0.01, 0.05, 0.1, and 0.2 to observe different spreading behaviours in Cooper's Instagram network, and we ran the model for 15 days to capture adoption efficiency.

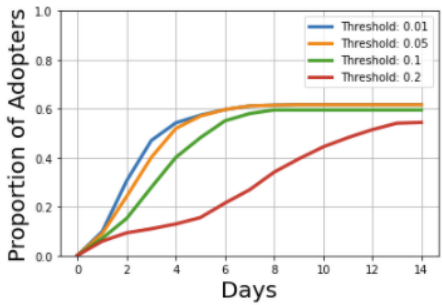

*Figure 2: Spreading behaviour with four threshold values*

To understand the spreading behaviour in the network, we decided to examine node 20 as it has the highest number of values in the centrality measures. Figure 2 shows the spreading behaviour of the node 20 with four threshold values. The proportion of adopters increases as the threshold value decreases, as seen in this graph, but there is a limit. When the threshold is lowered further, the proportion of adopters for node 20 does not increase any further than 0.6164. On the other hand, when the threshold value is raised, the proportion of adopters falls, and the number of days required to reach the maximum proportion increases. In the context of the cooper's instagram network, if users easily adopt a behaviour, the behaviour spreads much quickly and to many users. In addition to that, the node 20 reaches 66% of its proportion of adopter's value in just four days with the first three threshold values (0.01, 0.05 and 0.1). This means it requires fewer days for the majority of users to adopt a behaviour. The lower threshold values mean that followers do not need a significant amount of social influence to adopt a behaviour. However, this is not the case in real life, which is why, in the next section, we ranked nodes using the realistic threshold value (0.1).

### C. Ranking Nodes and Selecting the Optimal Node

| | node | in_degree | out_degree | eigenvector | betweenness | days_required | proportion_reached | score |
|---|---|---|---|---|---|---|---|---|
| 0 | 20 | 59 | 59 | 0.39108 | 0.17975 | 8 | 0.595082 | 7.438525 |
| 1 | 51 | 17 | 18 | 0.20250 | 0.01852 | 8 | 0.595082 | 7.438525 |
| 2 | 52 | 46 | 36 | 0.34206 | 0.09510 | 8 | 0.595082 | 7.438525 |
| 3 | 338 | 6 | 0 | 0.02888 | 0.00000 | 9 | 0.596721 | 6.630237 |
| 4 | 41 | 6 | 7 | 0.07473 | 0.00424 | 9 | 0.595082 | 6.612022 |
| 5 | 53 | 28 | 22 | 0.25556 | 0.04210 | 9 | 0.595082 | 6.612022 |
| 6 | 110 | 13 | 17 | 0.05888 | 0.04254 | 9 | 0.595082 | 6.612022 |
| 7 | 111 | 8 | 16 | 0.03769 | 0.01219 | 9 | 0.595082 | 6.612022 |
| 8 | 106 | 7 | 6 | 0.03551 | 0.00127 | 9 | 0.595082 | 6.612022 |
| 9 | 46 | 16 | 20 | 0.21539 | 0.00866 | 9 | 0.595082 | 6.612022 |
| 10 | 224 | 11 | 8 | 0.09255 | 0.00407 | 9 | 0.595082 | 6.612022 |
| 11 | 37 | 20 | 16 | 0.23278 | 0.01551 | 9 | 0.595082 | 6.612022 |
| 12 | 327 | 9 | 5 | 0.01714 | 0.00559 | 9 | 0.595082 | 6.612022 |
| 13 | 227 | 8 | 8 | 0.06991 | 0.01225 | 9 | 0.595082 | 6.612022 |
| 14 | 337 | 6 | 3 | 0.03401 | 0.00506 | 9 | 0.595082 | 6.612022 |

*Table 6: Node ranked according to the spreading capability (score)*

| | node | in_degree | out_degree | eigenvector | betweenness | days_required | proportion_reached | score |
|---|---|---|---|---|---|---|---|---|
| 0 | 314 | 4 | 1 | 0.00809 | 0.00192 | 11 | 0.637705 | 5.797317 |
| 1 | 136 | 2 | 1 | 0.00006 | 0.00004 | 13 | 0.637705 | 4.905422 |

*Table 7: Node ranked according to proportion reached*

The column score in table 6 represents a node's spreading capability based on the equation (1.0) given above. The table is sorted in descending order by spreading capability (score), with the topmost node being the most influential node in the network. The column "days_required" denotes the efficiency with which the behaviour can be adopted across a network originating from a node, while the column "proportion reached" denotes the percentage of network users who adopted the behaviour. As can be seen, node 20 can cover nearly 60% of the network's central core in just 8 days, similar to the node 51 and 52, whereas other nodes in the table can also reach the same proportion but take more days. In addition, node 20 has the highest centrality measures (in degree, betweenness, and eigenvector), indicating that it is the most central node in the network and is close to other important nodes. Based on the data in table 7, the highest proportion of users who can visit Cooper's shop and sample their food is almost 63 percent, or approximately 384 users out of a total of 610 users present in the network's central core. Therefore, approximately 44% of users in the whole network can be the potential customers for Coopers.

| | node | in_degree | out_degree | eigenvector | betweenness | days_required | proportion_reached | score |
|---|---|---|---|---|---|---|---|---|
| node | 1.000000 | -0.382813 | -0.347528 | -0.322140 | -0.217458 | -0.316172 | -0.342790 | -0.373934 |
| in_degree | -0.382813 | 1.000000 | 0.868658 | 0.824600 | 0.842302 | 0.366156 | 0.447161 | 0.537754 |
| out_degree | -0.347528 | 0.868658 | 1.000000 | 0.757251 | 0.797759 | 0.275537 | 0.359418 | 0.442638 |
| eigenvector | -0.322140 | 0.824600 | 0.757251 | 1.000000 | 0.649520 | 0.222861 | 0.329641 | 0.421845 |
| betweenness | -0.217458 | 0.842302 | 0.797759 | 0.649520 | 1.000000 | 0.154151 | 0.223313 | 0.302460 |
| days_required | -0.316172 | 0.366156 | 0.275537 | 0.222861 | 0.154151 | 1.000000 | 0.968038 | 0.917749 |
| proportion_reached | -0.342790 | 0.447161 | 0.359418 | 0.329641 | 0.223313 | 0.968038 | 1.000000 | 0.981729 |
| score | -0.373934 | 0.537754 | 0.442638 | 0.421845 | 0.302460 | 0.917749 | 0.981729 | 1.000000 |

*Table 8: Correlation between centrality measures and spreading capability*

The correlation between different measures is shown in table 8. It shows that the number of indegrees (followers) is more correlated with the proportion of adopters and the days required to reach the maximum proportion, followed by the eigenvector centrality, and finally the betweenness centrality. In other words, the betweenness centrality gives very little information as compared to other centrality measures about the spreading capability of a node.

According to table x.2, the node 314 reaches the maximum proportion of users which is nearly 63% of the network's central core and it has the highest value among its peers. However, it has less spreading capability since it is a few hops away from influencers which leads to additional days. The node 20, even though it reaches nearly 60% of the central nodes, takes fewer days as compared to the node 314. Moreover, the node 314 has a fewer number of in-degrees, which makes it a normal user, rather than an influencer. The table also shows that nodes 51 and 52 have the same spreading capability as node 20. However, if we take all the centrality measures into consideration, we find that node 20 has the highest value for all the measures.

Therefore, we recommend node 20, which is the most central and most important node in the network, as the influencer for the Coopers' marketing strategy program.

## VI. Conclusion

The advent of social media has allowed businesses to market their products to select their target audience. Influencer marketing is a relatively new but highly successful avenue for businesses to promote their products. This paper analysed the use of Social Network Analysis to detect the ideal instagram influencer for a company named Veganwich. Concepts such as centrality measures and cascading models were taken into consideration as part of the analysis process. Centrality measures are common methods used for identifying the most central node in the network, which could help in disseminating information in the network faster, stopping the epidemics and protecting the network from breaking. Whereas the cascading models assists in understanding the spreading behaviour of a group of nodes.

Every centrality measure has their strengths, however, from our analysis and derived values, we identified various drawbacks which led us to conclude that centrality measures on their own are not enough to identify influencers. Betweenness is a great tool for finding users who can spread information quickly in the network in terms of spreading information from one individual to another. However, it fails to identify whether the users are influencers themselves. On the other hand, degree centrality helps us to identify users with the most followers but fades away when trying to understand the spread of information among the followers of the influencers. Eigenvector allowed us to identify nodes with influential connections. However, it fails to show us the efficiency of the spread of information. Therefore, to understand the information spreading capability of each user, a linear threshold model was used to simulate the behaviour spreading in the network. Thus, we considered the correlation between different centrality measures and the spreading capability of users to develop our final recommendation.

From our analysis and data provided above, we recommend user 20 as the influencer for the Coopers' marketing strategy plan. This user is the most central, disseminates the information faster and is situated near other influential users as shown by the centrality measures. In addition to that, the linear threshold model shows the user 20 can spread the behaviour faster and to a higher number of users in the network. In conclusion, the combined result of the threshold model and centrality measure proves that user 20 is the most optimal influencer in the network to whom the Cooper's should offer sponsorship.

[15] "Do influencers need to tell audiences they're getting paid?," Harvard business review, 29-Aug-2019.